\documentclass[aps,twocolumn,superscriptaddress,nofootinbib]{revtex4-2}
\usepackage{amsmath,amssymb,graphicx}
\usepackage{hyperref}
\usepackage{bm}
\usepackage[normalem]{ulem}
\usepackage{microtype,amsfonts,amstext,mathtools,physics}
\hypersetup{colorlinks=true,linkcolor=blue,citecolor=blue,urlcolor=blue}

\begin{document}

\title{On the Cosmological Constant in Pre-geometric Gravity}

\author{Andrea Addazi}
	\email{andrea.addazi@qq.com}
	\affiliation{School of Physics and Astronomy, Anqing Normal University, Anqing 246133, People's Republic of China}
	\affiliation{Institute of Astronomy and Astrophysics, School of Mathematics and Physics, Anqing Normal University, Anqing 246133, China}
	\affiliation{CAS Key Laboratory for Research in Galaxies and Cosmology, School of Astronomy and Space Science, University of Science and Technology of China, Hefei 230026, China}
	\affiliation{Laboratori Nazionali di Frascati INFN, Frascati (Rome), Italy, EU}

\author{Giuseppe Meluccio}
	\email{giuseppe.meluccio-ssm@unina.it}
	\affiliation{Scuola Superiore Meridionale, Largo San Marcellino 10, Napoli 80138, Italy}
	\affiliation{INFN Sezione di Napoli, Complesso Universitario di Monte Sant'Angelo, Edificio 6, Via Cintia 21, Napoli 80126, Italy}
	\affiliation{Department of Physics, Loyola Marymount University, 1 LMU Drive, Los Angeles 90045, California, USA}

\date{\today}

\begin{abstract}
We offer novel perspectives on the cosmological constant (CC) problem, aiming toward a potential solution that requires neither fine-tuning nor anthropic arguments. In pre-geometric gravity (PGG), spacetime emerges from the spontaneous breaking of a fundamental gauge symmetry. This mechanism dynamically generates general relativity while also revealing a deep connection: the topological Gauss--Bonnet coupling of the theory scales precisely as the de Sitter entropy, an enormous number which reflects the information content of our universe. This coupling acts as a gravitational $\theta$-angle parameter, forcing the CC to become quantized into discrete topological sectors. The symmetry-breaking dynamics naturally selects the sector corresponding to the observed vacuum energy. The selected vacuum state is stabilized by the extremely large potential barrier of the pre-geometric Higgs field, which effectively seals it off from quantum tunneling transitions to other topological sectors. The PGG framework could thus provide a dynamical explanation for the smallness of the CC, linking gravity, topology and quantum information in a unified picture.
\end{abstract}

\maketitle

\section{Introduction}
The cosmological constant (CC) problem embodies the most profound failure of effective field theory reasoning in fundamental physics. Quantum fluctuations in the Standard Model vacuum contribute to an energy density $\rho_{\text{vac}} \sim M_{\text{EW}}^4 \sim (10^3\ \text{GeV})^4$, exceeding the observed dark energy density $\rho_{\Lambda} \sim (10^{-12}\ \text{GeV})^4$ by 60 orders of magnitude. Including Planck-scale physics exacerbates the discrepancy to about 120 orders of magnitude \cite{weinberg}. Equally perplexing is the fact that the electroweak scale remains inexplicably separated from the Planck mass $M_{\text{P}} \sim 10^{19}$ GeV despite receiving quadratically divergent radiative corrections.

These hierarchies share a striking numerical coincidence: the de Sitter (dS) entropy $S_{\text{dS}} = 3\pi/(G\Lambda) \sim 10^{120}$ precisely mirrors the ratio $M_{\text{P}}^2/\Lambda$ between the Planck scale $M_{\text{P}}\equiv(8\pi G)^{-1/2}$, where $G$ is the gravitational constant, and the cosmological constant $\Lambda$ \cite{Addazi:2020axm,Addazi:2020wnc,Addazi:2020mnm}. This suggests that information-theoretic principles -- specifically holographic entropy bounds -- may hold the key to understanding vacuum selection and stabilization. Yet no existing framework has so far managed to successfully convert this observation into a dynamical mechanism that simultaneously explains both hierarchies without the need of invoking anthropic selection arguments or fine-tuned parameters.

Here we present new arguments stemming from pre-geometric gravity (PGG), a framework in which spacetime geometry itself is not fundamental but rather emergent from the spontaneous symmetry breaking (SSB) of a gauge theory coupled to a Higgs-like field \cite{Addazi:2024rzo,Addazi:2025vbw,Addazi:2025qkc,Meluccio:2025uyo,Addazi:2025gcl,addazi:emergence,capozziello:pre-geometric,meluccio:black-holes}. Our central result is that the MacDowell--Mansouri (MM) action for the de Sitter group $SO(1,4)$ forces the Gauss--Bonnet (GB) coupling of the emergent gravitational theory to equal the de Sitter entropy up to a numerical factor. This identification transforms the cosmological constant from a continuous parameter into a topologically quantized number, determined by an integer $k$.

Unlike previous approaches that treat the CC as a variable to be selected from a landscape of possibilities, our mechanism provides a deterministic vacuum selection: the symmetry-breaking potential of the theory possesses an infinite number of degenerate minima, each corresponding to a distinct value of $\Lambda$. The observed vacuum state corresponds to the unique sector selected by the spontaneous symmetry breaking scale $v \sim 10^{120}$ (in natural units), with transitions between sectors suppressed by an instanton barrier of height $e^{-S_{\text{dS}}}$ or $e^{-v}$. This realizes the $\mathcal{H}$olographic $\mathcal{N}$aturalness program \cite{Addazi:2020axm,Addazi:2020wnc,Addazi:2020mnm}, which leverages the insight that a transition from a state with a relatively small CC (large entropy $S_{\text{dS}} \sim N \gg 1$, where $N$ is the number of holographic qubits) to a state with a Planckian CC (unit entropy) is not merely a loop correction, but a quantum process that must overcome an exceptionally high entropic barrier. The amplitude for such a decay is exponentially suppressed as $e^{-N}$, thus inherently protecting the relatively small value of the CC. From this point of view, the standard quantum field theory computations of vacuum bubbles are incomplete for they fail to account for the thermal, information-rich environment of the de Sitter horizon.

The paper is organized as follows. Section II reviews the pre-geometric framework and demonstrates how the Gauss--Bonnet coupling inevitably inherits the right entropy scale. In Section III we establish the topological origin of the cosmological constant quantization and derive the selection of the $k \sim 10^{120}$ sector, other than analyze the stability of the selected vacuum against quantum tunneling and radiative corrections. Section IV presents conclusions and remarks.

\section{Spontaneous Symmetry Breaking and Emergent Geometry}
We consider a gauge theory defined on a four-dimensional manifold that initially lacks any metric structure. The gauge group is taken to be the de Sitter group \(SO(1,4)\), with connection \(A_{\mu}^{AB}\) and curvature \(F_{\mu\nu}^{AB}\). The theory is required to be generally covariant, but neither a spacetime metric nor tetrads are introduced at the fundamental level. The only fixed tensorial structure on each tangent space is an internal metric \(\eta\) with signature \((-,+,+,+,+)\), which generalizes the Minkowski metric.

The spacetime geometry emerges dynamically through the condensation of a scalar field \(\phi^{A}\) transforming under the fundamental representation of \(SO(1,4)\). Via an adequate SSB potential for $\phi$, this gravitational Higgs field can acquire a nonzero vacuum expectation value (VEV) \(\langle \phi \rangle\) which spontaneously breaks the gauge symmetry down to that of the Lorentz subgroup \(SO(1,3)\). In the broken phase, the low-energy effective theory reproduces general relativity, with the pre-geometric fields \(A_{\mu}^{AB}\) and \(\phi^{A}\) turning effectively into gravitational degrees of freedom.

Since no inverse metric exists before the SSB, the only available contravariant density on the manifold is the Levi-Civita symbol \(\epsilon^{\mu\nu\rho\sigma}\), which has weight \(-1\). This uniquely fixes the form of generally covariant Lagrangian densities. The MacDowell--Mansouri Lagrangian \cite{macdowell:unified} is one such possibility:
\begin{equation}
\mathcal{O}_{\text{MM}} = Y_{\text{MM}} \, \epsilon_{ABCDE} \, \epsilon^{\mu\nu\rho\sigma} \, F_{\mu\nu}^{AB} F_{\rho\sigma}^{CD} \, \phi^{E},
\label{eq:MM_action}
\end{equation}
where uppercase Latin indices \(A,B,\dots\) run from 1 to 5, Greek indices denote spacetime coordinates and \(Y_{\text{MM}}\) is a coupling constant of mass dimensions \([\phi]^{-1}\).

After the SSB produced by a `sombrero' potential for $\phi$, one internal direction is fixed by the VEV \(\langle \phi^{A} \rangle = v \delta^{A}_{5}\) \cite{Addazi:2024rzo}. The \(SO(1,4)\) connection then decomposes as
\[
A_{\mu}^{a5} \quad (\text{four components}), \qquad
A_{\mu}^{ab} \quad (\text{six components}),
\]
with \(a,b = 1,\dots,4\) denoting Lorentz indices. By identifying
\begin{equation}
e_{\mu}^{a} \equiv m^{-1} A_{\mu}^{a5}, \qquad
\omega_{\mu}^{ab} \equiv A_{\mu}^{ab},
\label{eq:identification}
\end{equation}
where \(e_{\mu}^{a}\) are tetrads, \(\omega_{\mu}^{ab}\) the spin connection and \(m\) a mass scale, the framework reduces to a gravitational theory with an emergent pseudo-Riemannian geometry. Indeed, substituting the VEV of \(\phi^{A}\) into the MM Lagrangian yields
\begin{align}
	\begin{split}
		\label{eq:MM_SSB}
		\mathcal{O}_{\text{MM}} \;\xrightarrow{\text{SSB}}\; & 
		16 Y_{\text{MM}} v m^{2} \, e \, e_{a}^{\mu} e_{b}^{\nu} R_{\mu\nu}^{ab} \\
		&- 96 Y_{\text{MM}} v m^{4} \, e 
		- 4 Y_{\text{MM}} v \, e \, \mathcal{G},
	\end{split}
\end{align}
where \(e = \det e_{\mu}^{a}\), \(R_{\mu\nu}^{ab}\) is the curvature two-form of \(\omega_{\mu}^{ab}\) and \(\mathcal{G}\) is the Gauss--Bonnet density. These three terms correspond, respectively, to the Einstein--Hilbert action, the cosmological constant term and a topological invariant.

Matching the coefficients to the standard gravitational action gives the emergent Planck mass
\begin{equation}
M_{\text{P}}^{2} = 32 Y_{\text{MM}} v m^{2}
\label{eq:planck_mass}
\end{equation}
and the emergent cosmological constant
\begin{equation}
\Lambda = 3 m^{2} = \frac{3 M_{\text{P}}^{2}}{32 Y_{\text{MM}} v}.
\label{eq:cosmo_constant}
\end{equation}
For \(k_{\text{MM}} \sim \mathcal{O}(1)\) and the observed value \(M_{\text{P}}^{2} \sim 10^{37}\,\text{GeV}^{2}\), the relatively small measured value \(\Lambda \sim 10^{-84}\,\text{GeV}^{2}\) follows from a relatively large VEV \(v \sim 10^{120}\,[\phi]\).

An important observation is that the Gauss--Bonnet coupling \(\alpha_{\text{GB}} = -4 Y_{\text{MM}} v\) is not independent: using Eqs.\ \eqref{eq:planck_mass} and \eqref{eq:cosmo_constant} leads to the universal relation 
\begin{equation}
\alpha_{\text{GB}} = - 4 Y_{\text{MM}} v = - \frac{3 M_{\text{P}}^{2}}{8\Lambda} \, .
\label{eq:GB_relation}
\end{equation}
For the observed value of the CC, this requires \(|\alpha_{\text{GB}}| \sim 10^{120}\), a number that scales as the de Sitter entropy of the observable universe:
\begin{equation}
\label{comm}
|\alpha_\text{GB}|\sim S_\text{dS} \sim \frac{R_\text{H}^2}{L_\text{P}^2}\, ,
\end{equation}
where $R_\text{H}\sim 1/\sqrt{\Lambda}$ is the Hubble radius. 
This unexpected link between the pre-geometric symmetry breaking, the smallness of \(\Lambda\) and holographic information will be central to our subsequent discussion of topological stabilization and protection.

As remarked in Ref.\ \cite{HNPGG}, the de Sitter entropy is also scaling as $Y_\text{MM}v$, which suggests that the pre-geometric gravitational Higgs field $\phi$ also plays the rule of an {\it information field}. Moreover, the largeness of the GB coupling is related to the largeness of the VEV scale $v$ of $\phi$ and the smallness of the CC as an {\it information see-saw mechanism}. It is worth noting that the GB coupling scales as the inverse of the gravitational coupling at the Hubble scale:
\begin{equation}
\label{GBG}
|\alpha_\text{GB}| \sim \frac{1}{\alpha_{G}(\Lambda)},\qquad \alpha_{G}(\Lambda)=\frac{\Lambda}{M_\text{P}^{2}}\, .
\end{equation}

In a broader sense, the theory under consideration is reminiscent of the electric-magnetic coupling duality found in the case of Dirac's monopole/string \cite{Dirac} (where the string defect can be eliminated as shown in \cite{WuYang}) and in P-brane dyons \cite{Deser}, an intriguing analogy with the Seidberg--Witten duality \cite{SW}. Similar dualities were already found in the context of higher dimensional (super)gravity and string theories \cite{Hull:2000zn,Hull:2023iny} while analyzing magnetic-like charges for gravitons and dualities. The analogy here can be pushed even further: the fact that the dS entropy is expected to count a finite natural number of information qubits holographically stored on the Hubble horizon can be related to the quantization of the GB charge as follows:
\begin{equation}
\label{quantt}
|\alpha_\text{GB}| \alpha_{G}(\Lambda) \sim 1 \rightarrow S_\text{dS}\sim |\alpha_\text{GB}| \sim \frac{1}{\alpha_G(\Lambda)}\sim N\, .
\end{equation}
As it is known, in $d=4$ spacetime dimensions the Gauss--Bonnet term is topological, being a surface integral which does not contribute to the field equations. In addition to that, whilst it is true that the GB term is non-topological for spacetime dimensions $d>4$, this option is automatically discarded in the PGG framework because of the way the MM operator \eqref{eq:MM_action} is constructed (in terms of the Levi-Civita symbol).

When integrated over a smooth, compact $4$-manifold $M$ without boundary, the GB term is proportional to a topological invariant as
\begin{equation}
\label{MGB}
\frac{1}{32\pi^2}\int_M e \mathcal{G} \, d^4x = \chi(M)\, ,
\end{equation}
where $\chi(M)$ is the Euler characteristic of the manifold. The topological invariant  $\chi(M)$ is an integer for compact manifolds; for example, $\chi(S^4) = 2$ (the Euclidean dS instanton), $\chi(T^4) = 0$, $\chi(\mathbb{CP}^2) = 3$, etc. In quantum gravity, Eq.\ \eqref{MGB} typically receives contributions from instantons and $\chi(M)$ is related to the topological winding number. In the gravitational path integral, different topologies will then contribute with a weight given by
\begin{equation}
\label{EqP}
\begin{split}
	\Psi&=\int \mathcal{D}A \mathcal{D}\phi e^{iS[A,\phi]}\\
	&\xrightarrow{SSB} \int \mathcal{D}g \mathcal{D}\omega e^{iS_\text{EH}[g,\omega]}e^{32\pi^2 i\alpha_\text{GB} \chi(M)}\, ,
\end{split}
\end{equation} 
hence $\alpha_\text{GB}$ acts precisely as a $\theta$-angle for gravity:
\begin{equation}
\label{thetta}
\theta=32\pi^2 \alpha_\text{GB}\quad ({\rm mod}\,2\pi)\, .
\end{equation}

\section{Quantization Condition for the Cosmological Constant}
Eq.\ \eqref{EqP} provides a periodicity condition which leads to the quantization law of $\theta$. Moreover, the $\theta$-angle, from Eqs.\ \eqref{eq:GB_relation} and \eqref{thetta}, is related to the PGG Higgs VEV as
\begin{equation}
\label{thetaa}
\theta=-c Y_\text{MM}v
\end{equation}
with $c= 128\pi^2$. Indeed, redefining $\theta$ with an additional factor $2\pi k$ with $k\in\mathbb{Z}$ provides the same physical contribution to the gravitational path integral, as
\begin{equation}
	\label{periodic-theta}
	\theta\rightarrow \theta+2\pi k\quad\Rightarrow\quad e^{i\theta\chi(M)}\rightarrow e^{i\theta\chi(M)+2\pi ik\chi(M)}=e^{i\theta\chi(M)}
\end{equation}
for any value of the Euler characteristic. From a physical perspective, $k$ distinguishes between the different topological sectors of the theory. Such a discretization has analogies with the quantum Hall effect, as recently pointed out in Ref.\ \cite{Alexander:2025qkx}.

In the context of PGG, this result for $\theta$ can be corresponded with a possible periodicity of the  pre-geometric Higgs field $\phi$, which can possess a large number of equivalent vacuum states since
\begin{equation}
\label{periodic}
\theta\rightarrow \theta+2\pi k\quad\Rightarrow\quad- c Y_\text{MM}v\rightarrow- c Y_\text{MM}v+2\pi k\, .
\end{equation}
This condition cannot be satisfied by a typical `sombrero' potential, rather it demands a periodic potential for the gravitational Higgs field like
\begin{equation}
\label{laa}
\begin{split}
	V&=Y_\text{SSB} |J|\Big[1- \cos \Big(    cY_\text{MM}(|\phi|-v)\Big)\Big]^{p}\\
	&= Y_\text{SSB} |J|\Big[2\sin^{2}\Big(    \frac{c}{2}Y_\text{MM}(|\phi|-v)\Big)\Big]^{p}\, ,
\end{split}
\end{equation}
where $|\phi|=\sqrt{|\eta_{AB}\phi^{A}\phi^{B}|}$, $p$ is a positive semi-integer and
\begin{equation}
\label{J}
J=\epsilon_{ABCDE} \epsilon^{\mu\nu\rho\sigma} \nabla_\mu \phi^A \nabla_\nu \phi^B \nabla_\rho \phi^C \nabla_\sigma \phi^D \phi^E
\end{equation}
is the pre-geometric effective volume form factor, given that $|J|\xrightarrow{SSB} v^5 m^4 e$. As a consequence, the initially allowed number of vacuum states is now orbifolded to $\mathbb{R}/\mathbb{Z}_k$. The case $p=1$ reminds of the same well-known structure from instanton-generated potentials for QCD axions \cite{Axion1,Axion2,Axion3}. The relation between the CC and $\mathbb{Z}_N$ was also envisaged in Ref.\ \cite{HNPGG} from another prospective. Incidentally, let us also note that, by adding a pre-geometric Lagrange multiplier of the form $\lambda(J-\bar{J})$ with $\bar{J}$ constant, the potential \eqref{laa} would exactly reduce to a typical axion-like-particle potential except for characteristic energy scales.

More explicitly, the potential \eqref{laa} possesses a tower of discrete minima $v^{(0)}$, $v^{(1)}=v^{(0)}+2\pi/(c Y_\text{MM})$, etc., i.e.\ $v^{(k)}=v^{(0)}+2\pi k/(c Y_\text{MM})$. Using Eqs.\ \eqref{eq:GB_relation} and \eqref{thetaa}, the value of $v^{(0)}$ can be defined via the condition
\begin{equation}
\label{EqYone}
-c Y_{\text{MM}} v^{(0)} = -\frac{3c M_{\text{P}}^{2}}{32\Lambda^{(0)}} = \theta^{(0)}\, , 
\end{equation}
where $\theta^{(0)}\in[-\pi,0)$. The value $\theta^{(0)}=0$ is to be excluded on the grounds that it would correspond to the unbroken phase of the PGG framework (as $v^{(0)}=0$). Taken at face value, such an equation is satisfied if and only if $\Lambda^{(0)} \sim M_{\text{P}}^{2}$, which translates into an unreasonable Planckian value for the vacuum energy.

Combining Eq.\ \eqref{EqYone} with the periodicity condition \eqref{periodic} leads to a master equation for the generic $k$-th level of the tower of minima in terms of the $0$-th one:
\begin{equation}
\label{defff}
- c Y_\text{MM} v^{(k)} = - c Y_\text{MM} \Big(v^{(0)} + \frac{2\pi k}{c Y_\text{MM}}\Big)= \theta^{(0)} - 2\pi k\, ,
\end{equation}
thus yielding
\begin{equation}
\label{KeY!}
 \frac{3 c M_{\text{P}}^{2}}{32\Lambda^{(k)}} =- \theta^{(0)} + 2\pi k\, .
\end{equation}
This final relation is profound: the hierarchy between the Planck scale and the cosmological constant is controlled by the topological number $k$. A large value of $k$ leads to a large hierarchy between the two gravitational scales; in particular, for $k\gg1$ we obtain from Eq.\ \eqref{KeY!} that $M_{\text{P}}^{2}/ \Lambda^{(k)} \sim k \gg1$. The observed hierarchy between the two fundamental ultraviolet and infrared scales of our universe is then obtained for a value of $k$ which scales as the dS entropy:
\begin{equation}
\label{KN}
k \sim N \sim S_\text{dS} \sim \frac{M_{\text{P}}^{2} }{\Lambda}\, ,
\end{equation}
where the crucial identification  $\Lambda \equiv \Lambda^{(N)}$ was made in the last step. Therefore, a realistic cosmology arises when it is precisely the topological sector $k\sim N$ that is selected by the SSB mechanism of the pre-geometric theory.

The selection of this large topological sector follows from the principle of maximal entropy. The equilibrium probability for a given topological sector is $P_k \propto e^{S_{\text{dS}}(k)} \sim e^k$, hence the state with minimal $\Lambda$ (maximal $k$) is by far the most likely one. In the limit $\Lambda \to 0$ ($k \to \infty$), the de Sitter entropy diverges, making a configuration with vanishing cosmological constant infinitely more likely than any other state with larger $\Lambda$; in the limit $\Lambda \to \infty$, instead, the entropy vanishes. Therefore, the relaxation dynamics induced by the SSB naturally drives the system towards larger values of $k$ (smaller $\Lambda$), thus the observed smallness of $\Lambda$ emerges as a consequence of entropic naturalness rather than fine-tuning.

The present discussion also entails that the wave function of the universe corresponds to a vacuum state of maximal topological complexity. In other words, the CC vacuum is in a state of maximal topological complexity labeled by the quantum number $k$, as previously envisaged in Ref.\ \cite{Addazi:2020mnm}. Thus, the SSB of PGG selects a precise vacuum eigenstate as
\begin{equation}
\label{eigens}
|\Psi_{\theta}\rangle=\sum_{k\in\mathbb{Z}}c_{k}e^{i\theta_k}|k\rangle \xrightarrow{SSB} |N\rangle\, ,
\end{equation}
where, for every $k$, the $\theta_k$-angle is the phase corresponding to the specific vacuum state $|k\rangle$ and $c_{k}$ is a real coefficient.

A transition from the vacuum state $|N\rangle$ to $|N-1\rangle$ or $|N+1\rangle$ is suppressed by the tunneling barrier of the gravitational Higgs potential \eqref{laa}. Such a barrier is extremely large: considering for simplicity the case $p=1$ and expanding the potential around $v$ as $\phi^5=v+\rho$, we find that the quadratic mass term for $\rho$ scales as
\begin{equation}
\label{Massrho}
\mu_{\rho}^{2} \sim Y_\text{SSB} Y_\text{MM}^{2} m^{4}v^{5}\, ,
\end{equation}
which is, up to a free parameter $Y_\text{SSB}$, naturally a \mbox{(super-)Planckian} mass scale. This renders the potential barrier Planckian-stiff, implying a large exponential suppression of the tunneling probability from the CC vacuum state $|N\rangle$ to any other.

The excitation $\rho$ of the Higgs-like field $\phi^5$ post-SSB is expected to have a Planckian mass scale \cite{Addazi:2024rzo} in order to provide a possible ultraviolet completion of Einstein gravity, i.e.\ $\mu_{\rho}^{2} \sim M_\text{P}^{2}$. Comparing with Eqs.\ \eqref{eq:planck_mass} and \eqref{Massrho}, we thus obtain that
$$Y_\text{SSB} Y_\text{MM} m^{2}v^{4}\sim1\,,$$
which forces the value of the parameter $Y_\text{SSB}$ to be
\begin{equation}
\label{SSB}
Y_\text{SSB}\sim Y_\text{MM}^{-1}m^{-2}v^{-4} \sim  M_\text{P}^{-2}v^{-3}\, .
\end{equation}
If the $Y_\text{SSB}$ scale is associated to an effective mass scale $M_\text{SSB}$, this will then correspond to an energy which is much larger than the Planck scale:
\begin{equation}
\label{SSBB}
M_\text{SSB} \sim (M_\text{P}^{2}v^{3})^{1/5}\, . 
\end{equation}
In this scenario, once the CC is selected as $\Lambda \sim 10^{-120}\,M_{\text{P}}^{2}$, the potential barrier will protect it from any quantum tunneling processes. Furthermore, the fluctuation field $\rho$ will have a super-heavy mass related to the barrier curvature, which is thought of as Planckian; therefore, any radiative corrections coming from other coupled fields would perturb it only at energies such that the gauge symmetry $SO(1,4)$ is restored and the CC vanishes, as the pre-geometric spacetime would be restored.

On a final note, we observe that this stabilization mechanism is compatible with the entropic barrier picture envisaged from the $\mathcal{H}$olographic $\mathcal{N}$aturalness paradigm \cite{Addazi:2020axm,Addazi:2020wnc,Addazi:2020mnm}. Indeed, since the the VEV $v$ scales as the dS entropy, moving away from lower values of $v$ would also imply a spontaneous violation of the second law of thermodynamics: if the dark energy mass flowed to the Planck scale through radiative corrections, then entropy would change from the very large number $N\sim 10^{120}$ to an $O(1)$ value, a process which is probabilistically suppressed as
\begin{equation}
\label{Suppress}
|\langle N| 1\rangle|^2 \sim e^{-S_\text{dS}}\sim e^{-1/\alpha_{G}(\Lambda)}\sim e^{-N}\sim e^{-10^{120}}\, . 
\end{equation}
The $\alpha_{G}^{-1}$ scaling is typical of instantonic contributions, which are compatible with the pre-geometric potential suppression. This exponential suppression directly addresses the problem of the radiative instability of the CC. In our framework, indeed, $\Lambda$ is not a continuous parameter subject to perturbative renormalization but a topological quantum number tied to the de Sitter entropy via $S_{\text{dS}} \sim M_{\text{P}}^2/\Lambda \sim k$. Any quantum correction that would increase $\Lambda$ would also decrease the entropy by $\delta S \sim -S_{\text{dS}} (\delta\Lambda/\Lambda)$. Such a decrease violates the second law of thermodynamics and is exponentially suppressed with probability $e^{-S_{\text{dS}}} \sim e^{-10^{120}}$. Thus, radiative corrections that would make $\Lambda$ larger are effectively forbidden. Concerning variations of $\alpha_{\text{GB}}$, these are not expected from quantum radiative corrections coming from the bulk dynamics, as $\alpha_{\text{GB}}$ is a topological charge related to a boundary term. The relation between $\alpha_{\text{GB}}$ and $\Lambda$ is preserved under renormalization not by a symmetry but by the protection mechanism of holographic entropy.

\section{Conclusions}
We have presented a pre-geometric approach to quantum gravity that addresses the cosmological constant problem. Starting from an $SO(1,4)$ gauge symmetry and introducing spontaneous symmetry breaking via a gravitational Higgs field, an emergent gravitational theory is obtained, as the MacDowell--Mansouri action naturally generates the Einstein--Hilbert, the cosmological constant and the Gauss--Bonnet terms.

The key results of our analysis are the following.

\begin{itemize}
    \item \textbf{Emergent gravitational scales and entropic see-saw mechanism.} The SSB of the pre-geometric $SO(1,4)$ gauge group dynamically generates the Planck mass $M_{\text{P}}$ and the cosmological constant $\Lambda$. The theory thus reveals a fundamental see-saw relation, $M_{\text{P}}^2/\Lambda \sim v$, where the vacuum expectation value of the gravitational Higgs field $\phi$ is the relatively huge number $v \sim 10^{120}\,[\phi]$. Crucially, this ratio is shown to be proportional to the Gauss--Bonnet coupling $\alpha_{\text{GB}}$, which itself scales exactly as the de Sitter entropy $S_{\text{dS}} \sim M_{\text{P}}^2/\Lambda$. This provides a direct and dynamical link between the ultraviolet and the infrared scales of gravity by means of holographic information.

    \item \textbf{Topological quantization of the cosmological constant.} The Gauss--Bonnet term, which is a total derivative in four spacetime dimensions, is recognized as a topological $\theta$-term in the gravitational path integral. This forces its coupling $\alpha_{\text{GB}}$ -- and therefore $\Lambda$ via the see-saw relation -- to be quantized. The quantization condition $\Lambda \propto 1/(k G)$ ties the observed value of the cosmological constant directly to an integer topological quantum number $k$, thereby transforming a continuous fine-tuning problem into a mechanism of discrete vacuum selection via SSB.

    \item \textbf{Deterministic selection of our universe.} The SSB potential for the pre-geometric Higgs field is periodic, leading to a discrete landscape of topologically distinct vacua, each one corresponding to a different value of $k$ and thus of $\Lambda$. The selection of the large topological sector $k \sim 10^{120}$ follows from the principle of maximal entropy: the equilibrium probability $P_k \propto e^{S_{\text{dS}}(k)} \sim e^k$ overwhelmingly favors the state with minimal $\Lambda$. This fact reveals that the smallness of the observed CC is a direct outcome of entropic naturalness. Furthermore, our universe is identified as a state of maximal topological complexity.

    \item \textbf{Dynamical stabilization from the entropic barrier.} The selected vacuum state is dynamically protected against quantum tunneling to other topological sectors by an enormous entropic barrier. Indeed, the potential barrier between different vacua is Planckian and the transition probability is suppressed by a factor $e^{-S_{\text{dS}}} \sim e^{-10^{120}}$, effectively stabilizing the CC at all times. This process naturally solves the problem of the radiative instability of the CC: any quantum correction that would increase $\Lambda$ would also decrease entropy and is thus exponentially suppressed. Therefore, the relation between $\alpha_{\text{GB}}$ and $\Lambda$ is preserved under renormalization by the protection mechanism of holographic entropy. This process is a dynamical realization of the $\mathcal{H}$olographic $\mathcal{N}$aturalness paradigm and shows that the vacuum energy is protected not by a symmetry but by the exponentially suppressed probability of transitioning from a state of high entropy to one of lower entropy. Radiative corrections are also controlled in this approach, as they cannot alter the fundamentally topological and quantized nature of $\Lambda$.
\end{itemize}

Our framework offers several promising directions for future research. The pre-geometric perspective could shed light on the nature of quantum gravity near the Planck scale, where conventional geometric notions break down. The connection between gravitational topology and information theory corroborates the deep links between quantum gravity and quantum information. The stabilization mechanism for $\Lambda$ may have implications for inflationary cosmology as well as for the late-time acceleration of the universe. Moreover, higher-dimensional operators such as $\mathcal{O}_{\text{MM}}^2$ may generate a dynamical type of dark energy which is stable under quantum corrections \cite{InPrep}, providing a possible natural explanation for recent DESI observations \cite{DESI:2024mwx}.

\vspace{0.5cm}

{\bf Acknowledgments.}
The authors would like to thank Salvatore Capozziello and Yermek Aldabergenov for discussions and remarks on these subjects. AA's work was supported by the Program for Innovative Research Team in Anqing Normal University and the National Science Foundation of China (NSFC) through grant No.\ 12350410358. GM acknowledges the support of  Istituto Nazionale di Fisica Nucleare (INFN), Sezione  di Napoli, \textit{Iniziativa Specifica} QGSKY. This publication is based upon work from COST Action CA21136 -- ``Addressing observational tensions in cosmology with systematics and fundamental physics (CosmoVerse)'', supported by COST (European Cooperation in Science and Technology).

\end{document}